\documentclass[conference]{IEEEtran}
\IEEEoverridecommandlockouts
\usepackage{amsmath,amsfonts}

\usepackage{algorithm}
\usepackage{array}
\usepackage[caption=false,font=normalsize,labelfont=sf,textfont=sf]{subfig}
\usepackage{textcomp}
\usepackage{stfloats}
\usepackage{url}
\usepackage{verbatim}
\usepackage{graphicx}
\usepackage{cite}
\usepackage{tabularx}
\usepackage{makecell}
\usepackage[top=2cm, bottom=2cm, left=2cm, right=2cm]{geometry}
\usepackage{algorithm}
\usepackage{algorithmicx}
\usepackage{algpseudocode}
\usepackage{amsmath}
\usepackage{amsthm}
\usepackage{makecell}
\usepackage{url}
\usepackage{cite} 
\usepackage[numbers,sort&compress]{natbib}
\usepackage{graphicx}
\usepackage{bbding}
\usepackage{graphicx}
\usepackage{subfig}
\usepackage{overpic}
\usepackage{float} 

\usepackage[misc]{ifsym}
\def\BibTeX{{\rm B\kern-.05em{\sc i\kern-.025em b}\kern-.08em
    T\kern-.1667em\lower.7ex\hbox{E}\kern-.125emX}}

\begin{document}

% \title{Topic and Stance Aware Information Diffusion Model in Online Social Networks}
\title{Exploring the Independent Cascade Model and Its Evolution in Social Network Information Diffusion}
% \author{Xinyu Li, Chenwei Wang, Jinyang Huang, Xiao Sun, Meng Wang

\author{
	\IEEEauthorblockN{
		Jixuan He\IEEEauthorrefmark{1}, 
		Yutong Guo\IEEEauthorrefmark{1},
        Jiacheng Zhao\IEEEauthorrefmark{1} 
		} 
	\IEEEauthorblockA{\IEEEauthorrefmark{1}School of Computer and Information, Hefei University of Technology, Hefei, China.}

}

\maketitle

\begin{abstract}
This paper delves into the paramount significance of information dissemination within the dynamic realm of social networks. It underscores the pivotal role of information communication models in unraveling the intricacies of data propagation in the digital age. By shedding light on the profound influence of these models, it not only lays the groundwork for exploring various hierarchies and their manifestations but also serves as a catalyst for further research in this formidable field.

\end{abstract}


\section{Introduction}
In today's Internet, that is, the era of interconnection, a variety of online social network platforms have penetrated into all aspects of our lives, forming a very complete, comprehensive and convenient platform for information exchange. In today's society, this rich online ecosystem of digital connectivity not only defines new ways for people to communicate with each other, but also offers a wealth of personal stories and a growing global news feed.

For all kinds of Internet, sales types of companies and related marketers, it is important to accurately understand and anticipate how much the spread of positive and negative information on these online network platforms affects our daily lives. In the context of epidemics and emergency response, the ability to disseminate vital information quickly can be a matter of life and death.

In this paper, we have discussed in depth the complexity of transmission models of various data types involved in online social platforms and the wide range of fields involved in these models, because these models are the basis for us to understand the transmission of various data in public social platforms, and they serve as tools for us to analyze information transmission and the development of things, so that we can understand the user locking mechanism.

As we begin this journey, the purpose of this article is to look to the future. These existing models play an integral role in data from the basis of communication models to practical applications, which also emphasizes the importance of these models. It's a way to understand the dynamics of information dissemination that not only reflects today, but also the new social media landscape.

It determines the key role of information and communication methods in the modern digital age. It highlights the overall concept of information in social networks and creates more research opportunities in a growing field.

With the stage set, our focus turns to the Independent Cascade Model (IC) and its evolution. The IC is an archetypal model for simulating the spread of information within social networks, serving as a cornerstone for numerous studies in the field. This is a model that simulates the spread of information in social networks and is the basis of much research in this field \cite{9484767}. Passing information from one person to another.

However, as the current online digital environment continues to evolve \cite{10339891}, it has become clear that the basic assumptions and limitations of ics may not fully understand the complexity of transmitting data of all kinds in the real world. This knowledge is the driving force behind our exploration and development of integrated circuits and evolutionary models, the limitations of which we attempt to address. By taking into account factors like time dynamics, relationships, and personal preferences.

To sum up, in this article, we not only introduce the broader context of the various information exchanges of online social networking platforms, but also lay the foundation for exploring the model and independent development of online social networking platforms.

\section{Related Work}
Kempe and his colleagues laid the cornerstone of information diffusion modeling by presenting two pivotal concepts: the Independent Cascade Model (IC)\cite{kempe2005influential} and the Linear Threshold Model (LT)\cite{kempe2003maximizing, 10443215}. These models, initially developed in the context of early social networks, offered valuable insights into how information spreads among connected individuals.

With the development of various digital landscapes in today's society and the exponential growth of online social network platforms becoming more and more obvious, the development of these platforms is getting faster and faster, and the number of participants is also showing exponential growth, which also leads to the spread of various false information, the spread of rumors, and even the occurrence of some dangerous events, causing irreparable social impact. In addition to simple relationships, these factors have led to changes in the trajectory of information transmission, including complex interactions as well as evolution in the transmission process. With the consideration of time, space, user emotion and personal preference, this broader view has been widely accepted by many scholars in contemporary online social network platforms, which has also prompted many researchers to explore the complexity of these dynamic processes of information transmission and explore how changing information patterns interact.

In the current field of online social platforms, many researchers have recognized the complex factors of mixed information transmission and the diversity of transmission processes in such platforms. It is these factors and the results that have led to the development of numerous innovative models, in which special attention is paid to the time dimension to improve the predictive accuracy of data transmission in an ever-changing digital environment.

For instance, in the quest to capture the dynamic nature of information propagation, Guille et al. introduced the T-BaSIC model\cite{guille2012predictive,10251628}. This pioneering model provides a comprehensive framework for understanding how information evolves and disseminates through social networks over time.

Haldar et al. took a unique approach by unveiling the Temporal Independent Cascade model (T-IC)\cite{haldar2023temporal,10149418}. This newly developed model is specifically designed to provide insight into the specific dependence and influence of time on the dissemination of various types of information across various online platforms, exploring in detail the impact of time spent on thin wear.

Chen and his team expanded the horizon of information propagation modeling by introducing the Independent Cascade Model with Meeting Events (IC-M)\cite{chen2012time}. This model has greatly exceeded the original standard independent cascade model, because in this new model, not only the interaction between users, but also the consideration of a wide variety of different events over a long period of time, this new finding provides a more detailed view of the study of information transmission in online social platforms.

Kim et al. added a distinctive perspective to the mix by enhancing the IC Model with their CT-IC model\cite{kim2014ct}, which brings continuous activation and time constraints into the fold\cite{9613773}. Meanwhile, Li et al. introduced the GT propagation model\cite{li2013modeling}, which extends the understanding of information propagation by considering the influence of time on user behavior. These models collectively contribute to a deeper understanding of the complex dynamics of information diffusion in the dynamic landscape of online social networks.

Exploring the intricate interplay between topics and information propagation, researchers have made significant strides in understanding how content disseminates in the realm of social networks. 

In their work\cite{barbieri2013topic}, they not only refined the method for assessing infection probabilities between users but also introduced an innovative topic-aware influence-driven propagation model. This model takes into account the thematic aspects of information diffusion, shedding light on the role of topics in shaping the spread of content.

Another noteworthy contribution comes from \cite{qin2021influence}, who introduced a topic-aware community-based independent cascade model. This approach dives deeper into the structural characteristics of communities while considering thematic features, providing a holistic view of information propagation within social networks.

In the pursuit of modeling the nuanced dynamics of social influence, \cite{tian2020deep} presented a topic-aware social influence propagation model. This model not only considers user-to-user interactions but also incorporates the thematic dimension, offering a more comprehensive understanding of information diffusion. Inspired by the PageRank algorithm, \cite{michelle2016topic} devised the TS-IDM model. This model brings topic sensitivity to the forefront, enabling a more nuanced assessment of information diffusion and its topic-related variations. \cite{zhang2020nsti} introduced the NSTI-IC model, a propagation model that takes into account non-adjacent users, enriching our understanding of the spread of information in social networks. Furthermore, \cite{chen2017modeling} ventured into spatial considerations by introducing a spatial Markov-dependent influence propagation model. The model is based on established integrated circuit models, but also incorporates spatial dynamics, a combination that reveals how geographic factors affect information dissemination in online platforms. Together, these pioneering efforts enhance our grasp of how topics and thematic characteristics affect the complex landscape of information diffusion in online web platforms.

Acknowledging the profound influence of user emotions on the intricate dynamics of information diffusion, \cite{wang2016emotion} introduces an emotion-driven independent cascade model. This model emphasizes the pivotal role emotions play in the intricate process of information spread within social networks. Furthermore, within the domain of emotional contagion during information dissemination, \cite{wang2017emotion} puts forward a sentiment-based independent cascade model. The proposal of this new and innovative model deeply studies the complex dynamics and extent of the influence of emotions on changes in the process of various types of information transmission, and this discovery reveals how emotions affect the trajectory of information transmission. Importantly, this model goes beyond traditional models of information dissemination to emphasize its efficacy in revealing and exploiting the emotions of information dissemination.

\section{Problem Formulation}
\subsection{Information Dissemination}
Various, even false information dissemination in online network platforms refers to the process of how information is spread from one user to another on online and online social platforms. This process can cover a variety of content, including news, opinions, pictures, videos, and links. The speed of information dissemination in online social networks is usually very rapid and extensive, which is related to many factors, not only related to personal interests, amateur platforms to detect the personality and other factors, therefore, in such platforms. Users can quickly share information and spread it to their followers and friends.

Information dissemination on social platforms usually involves a series of interactions between users, including the author, receiver, disseminator of the original information and the information sharing involved in the communication process. This process can be measured and tracked through a variety of interactions such as likes, likes, blocks, reports, comments, and shares.

The dissemination of information on social networks has a very important value, both social and commercial. First, commercial value means that merchants can spread their commercials and other messages in different ways on social media, with the goal of promoting a superior brand at once. Information communication can be used to promote products and services, disseminate information, promote discussion and promote social change. Therefore, it is very important for social media marketing to analyze public opinion and social science research, explore the models and laws of information dissemination on social networks and the influencing factors, including how these factors affect the communication process and how to control the influence of these factors.

\subsection{Independent Cascade Model}
This is an independent model published on an online social platform based on basic assumptions.

In the IC model, the model makes a preliminary assumption that information distributed on various online platforms can run freely across the entire network, that is, spread freely, and therefore can be added to snake wine. This assumption also means that on these platforms, the choice of users to receive and distribute their own behavioral data is separate from the behavior of other users, whether they want to manage or not, thus ensuring the independence of the data flow.

In probabilistic models: In IC models, users are faced with static probabilistic decisions when they receive data. It is important to know to send data to nearby users. The probability of these users sending data generally varies from user to user. This difference may depend on several factors, including relationships, personal interests, and other relevant considerations.

The integrated circuit model includes open data transfer at different time intervals at each stage, that is, information on the long-winded online platform, where users can decide to publish data to their network connection based on the data received, the information and the specific probability of data distribution.

Here are the detailed steps of how the IC model works:

Initial activation: The example first selects one or more seed points in the network. It was originally revived as a starting point for the dissemination of information.

Spread information: At each step, each activated seed node chooses to spread information to its neighbors based on its different spread possibilities. If the sending process is successful, the receive contract is activated and the message continues to be broadcast in the next steps.

Duration: Continue to repeat until any new contract is restarted, which indicates that the information has been completed or a predetermined time limit has been reached.

Track progress: Record active nodes and record specific active times throughout the broadcast.

The Importance of Linkage Mode (IC) is the independent framework that provides a simple and powerful simulation and understanding, and based on this information transfer mechanism, control of the social network platform online, the process is very similar to Kang logout, the model is dynamic rather than static. The researchers used integrated circuit models to assess the impact of variables on the propagation of information, such as the identification of affected nodes, possible changes in communication, and changes in network structure. The model has been widely used in HIV transmission detection, social marketing strategies and public opinion analysis.

\section{Disadvantages and Limitations}
The independent cascade model is a useful tool for promoting the spread of information in social networks, but as researchers continue to study it, they have found that there are some drawbacks and drawbacks to the model, including:

Independent assumption: The basic assumption of this model is that in each platform, no matter whether the information is true or false, or whether the information is transmitted by the user, the transmission of information is not affected by the behavior of other users, that is, in the summary of the network platform, each user's decision to accept and disseminate his own hot information is independent of the behavior of other users. However, real-world user decisions can be influenced by factors such as an individual's personality, interests, preferences, environment, interests, and personal beliefs. Therefore, these overly simple independent assumptions of this model may not reflect the complex dynamics of various information dissemination in the actual society.

Fixed probability: The independent cascade model relies on a fixed probability for each user, which is assumed to remain constant throughout the simulation. In fact, in these platforms, the user's behavior and potential to transmit information are very active and influenced by increasingly important factors such as social relationships, personality, environmental factors, interests, or recent experience, and cannot be static factors that are immutable. Therefore, assumptions about the possibility of transmission may not be sufficient to complicate the spread of information in an active social network. Thus, assumptions about the possibility of broadcasting may not be sufficient to complicate the spread of information in an active social network.

Node initiation threshold: The formal model does not include the concept of a user initiation threshold, which means that users may need to transmit information multiple times before deciding to propagate it further. This can be very important in some cases, but is not taken into account in the electronic model.

Network structure: In integrated circuit models, it is often assumed that the network structure is explicit and known in advance. However, the structure of social networks is often dynamic. These changes may come from the creation of a new connection, a current failure, or the emergence of a new network contract. This dynamic network structure is difficult to grasp and can lead to discrepancies between the model and the actual scenario.

Initial selection of contracts: In the example, the group of contracts to be initiated must be carefully selected, as this will have a significant impact on the outcome of the dissemination of information. During the launch process, the initial market selection may produce different results, which emphasizes the importance of a well-thought-out initial market selection.

\section{Model Evolution}
The evolution of the independent-level model involves strengthening and expanding the original model to better adapt to the actual dynamics of information dissemination in social networks. These developments include the introduction of new factors, expanded model applications, and more complex changes. The development trend of the independent level model is as follows:

\subsection{Temporal Evolution Models}
These models explain the changing nature of information over time, allowing the speed and patterns of information transmission to change and become more closely linked to real-world dynamics.

\subsection{Topic-aware models}
These models take into account the impact of content topics on the publishing process. They can choose different ways of disseminating information under different topics and reflect different levels of user interest in different topics.

\subsection{Emotion-aware models}
These models take into account the influence of user emotions on information dissemination. In particular, they explore in depth how users' emotional attitudes affect the reception and dissemination of various kinds of information in online platforms, and how to better understand the dissemination of emotional information.

\subsection{Spatial Awareness Models}
These models take into account how users' physical locations influence the spread of information. They specifically examine how the geographical distance between users impacts information transmission, making them suitable for scenarios where location plays a significant role in information propagation.

\subsection{Social relationship awareness models}
These models take into account the significance of social relationships among users in the context of information dissemination. They provide a more precise representation of the social network structure, encompassing relationships such as friends, followers, and fans.

\subsection{Dynamic communication models}
These models take into account the evolving nature of user behavior during the information transmission process. They can replicate shifts in user interests or the changing topology of social networks, offering a more realistic representation of how information propagates over time.

\subsection{Deep Learning Techniques}
The integration of deep learning approaches for modeling information propagation. The utilization of deep learning methods enhances the model's capability to handle vast datasets and intricate features, ultimately elevating predictive accuracy. Develop information dissemination models by integrating in-depth learning. The deep learning method improves the ability of the model to deal with large amounts of data and complex features, and finally improves the accuracy of the prediction.

The ongoing evolution of these models is driven by the pursuit of a more precise representation of information dissemination in social networks, catering to diverse application domains, and comprehensively accounting for the multifaceted factors influencing information transmission. These models are being further developed to more accurately reflect the dissemination of information in social networks, adjust different application areas, and take into account the multiple factors that influence the dissemination of information comprehensively. As social networks undergo continual transformation, researchers will persistently explore new avenues for enhancement and refinement, adapting to the evolving landscape of online interactions and information exchange.

\section{CONCLUSION}
This paper extensively explores the profound importance of information propagation in the ever-evolving landscape of social networks. It underscores the pivotal role played by information communication models in deciphering the complexities of data dissemination in the digital era. By illuminating the profound impact of these models, it not only establishes a solid foundation for investigating different hierarchies and their manifestations but also ignites the spark for continued research in this dynamic and influential field.



\bibliographystyle{unsrt}
\bibliography{con}
\end{document}